\documentclass[amsmath,amsfonts,prl,aps,twocolumn,twoside,superscriptaddress]{revtex4}

\usepackage{epsfig,verbatim}
\usepackage{graphicx}
\usepackage{amsmath}
\usepackage{revsymb}

\def\squareforqed{\hbox{\rlap{$\sqcap$}$\sqcup$}}

\def\qed{\ifmmode\squareforqed\else{\unskip\nobreak\hfil

\penalty50\hskip1em\null\nobreak\hfil\squareforqed

\parfillskip=0pt\finalhyphendemerits=0\endgraf}\fi}

\def\endenv{\ifmmode\;\else{\unskip\nobreak\hfil

\penalty50\hskip1em\null\nobreak\hfil\;

\parfillskip=0pt\finalhyphendemerits=0\endgraf}\fi}

\mathchardef\ordinarycolon\mathcode`\:

\mathcode`\:=\string"8000

\def\vcentcolon{\mathrel{\mathop\ordinarycolon}}

\begingroup \catcode`\:=\active

  \lowercase{\endgroup

  \let :\vcentcolon

  }

\newcommand{\nc}{\newcommand}

\nc{\rnc}{\renewcommand}

\nc{\lbar}[1]{\overline{#1}}

\nc{\bra}[1]{\langle#1|}

\nc{\ket}[1]{|#1\rangle}

\nc{\ketbra}[2]{|#1\rangle\!\langle#2|}

\nc{\braket}[2]{\langle#1|#2\rangle}

\nc{\proj}[1]{| #1\rangle\!\langle #1 |}

\nc{\avg}[1]{\langle#1\rangle}

\nc{\Rank}{\operatorname{Rank}}

\nc{\smfrac}[2]{\mbox{$\frac{#1}{#2}$}}

\nc{\Tr}{\operatorname{Tr}}

\nc{\tr}{\operatorname{Tr}}

\nc{\id}{\operatorname{id}}

\nc{\1}{\openone}

\nc{\ox}{\otimes}

\nc{\dg}{\dagger}

\nc{\dn}{\downarrow}

\nc{\cA}{{\cal A}}

\nc{\cB}{{\cal B}}

\nc{\cC}{{\cal C}}

\nc{\cD}{{\cal D}}

\nc{\cE}{{\cal E}}

\nc{\cF}{{\cal F}}

\nc{\cG}{{\cal G}}

\nc{\cH}{{\cal H}}

\nc{\cI}{{\cal I}}

\nc{\cJ}{{\cal J}}

\nc{\cK}{{\cal K}}

\nc{\cL}{{\cal L}}

\nc{\cM}{{\cal M}}

\nc{\cN}{{\cal N}}

\nc{\cO}{{\cal O}}

\nc{\cP}{{\cal P}}

\nc{\cQ}{{\cal Q}}

\nc{\cR}{{\cal R}}

\nc{\cS}{{\cal S}}

\nc{\cT}{{\cal T}}

\nc{\cX}{{\cal X}}

\nc{\cY}{{\cal Y}}

\nc{\cZ}{{\cal Z}}

\nc{\supp}{{\operatorname{supp}}}

\nc{\var}{\operatorname{var}}

\nc{\rar}{\rightarrow}

\nc{\lrar}{\longrightarrow}

\nc{\polylog}{\operatorname{polylog}}

\nc{\RR}{{{\mathbb R}}}

\nc{\CC}{{{\mathbb C}}}

\nc{\FF}{{{\mathbb F}}}

\nc{\NN}{{{\mathbb N}}}

\nc{\ZZ}{{{\mathbb Z}}}

\nc{\PP}{{{\mathbb P}}}

\nc{\QQ}{{{\mathbb Q}}}

\nc{\UU}{{{\mathbb U}}}

\nc{\EE}{{{\mathbb E}}}

\nc{\Icoh}{{I^{\rm coh}}}

\nc{\Qca}{{Q_{\rm ss}}}

\nc{\Qcaa}{{Q^{(1)}_{\rm ss}}}

\nc{\Dcaa}{{D^{(1)}_{{\rm ss}\rightarrow}}}

\nc{\Dca}{{D_{{\rm ss}\rightarrow}}}

\nc{\be}{\begin{equation}}

\nc{\ee}{{\end{equation}}}

\nc{\bea}{\begin{eqnarray}}

\nc{\eea}{\end{eqnarray}}

\nc{\<}{\langle}

\rnc{\>}{\rangle}

\nc{\Hom}[2]{\mbox{Hom}(\CC^{#1},\CC^{#2})}

\nc{\rU}{\mbox{U}}

\begin{document}

\title{Extensive Nonadditivity of Privacy}
\author{Graeme Smith}\email{gsbsmith@gmail.com}
\affiliation{IBM T.J. Watson Research Center, Yorktown Heights, NY
10598, USA}
\author{John A. Smolin}\email{smolin@watson.ibm.com}
\affiliation{IBM T.J. Watson Research Center, Yorktown Heights, NY
10598, USA}

\date{April 24, 2009}

\begin{abstract}
Quantum information theory establishes the ultimate limits on
communication and cryptography in terms of channel capacities for
various types of information.  The private capacity is particularly
important because it quantifies achievable rates of quantum key
distribution.  We study the power of quantum channels with limited
private capacity, focusing on channels that dephase in random bases.
These display extensive nonadditivity of private capacity: a channel
with $2\log d$ input qubits has a private capacity less than $2$,
but when used together with a second channel with zero private
capacity the joint capacity jumps to $(1/2)\log d$.  
In contrast to earlier work which found nonadditivity vanishing as a
fraction of input size or conditional on unproven mathematical
assumptions, this provides a natural setting manifesting nonadditivity of
privacy of the strongest possible sort.
\end{abstract}
\maketitle

{\em Introduction---}Communication channels are subject to interference 
and noise, even under the best operating conditions.  By modeling noise probabilistically, 
information theory characterizes the fundamental
limitations for communication in terms of the capacity of a channel \cite{Shannon48}.  The capacity,
measured in bits per channel use, establishes the boundary between communication rates that
are achievable in principle and those that aren't.  Furthermore, there is a simple formula for
the capacity, which can provide insight for designing practical protocols and give explicit bounds
on the performance of real-world systems \cite{RU03}. 

While a probabilistic description of noise is often a good approximation,
ultimately all communication systems are fundamentally quantum.  Furthermore,
in the regime where quantum effects become important there are several distinct 
notions of information transmission.  One may be interested in the capacity of a 
channel for classical, private, or quantum transmission.  The sender and receiver may
have access to some auxiliary resources, such as entanglement or classical communication.
The simplest case which involves no such assistance will be the focus of this letter.

The capacity of a channel for private classical communication \cite{D03} is of particular importance 
because of its relation to quantum key distribution \cite{BB84}.  The private capacity of a quantum channel $\cN$ is usually called $\cP(\cN)$, and 
is no larger than the classical capacity, $\cC(\cN)$.  Since fully quantum transmission is necessarily private, 
the private capacity of a channel is at least as large as its quantum capacity, $\cQ(\cN)$.  As a result, 
we have $\cQ(\cN) \leq \cP(\cN) \leq \cC(\cN)$.  In contrast to the classical capacity of a {\em classical} channel, no
simple expression is known for any of these three capacities of a quantum channel.  In fact, it is known that the 
natural guesses for $\cQ$, $\cP$, and $\cC$ are simply false \cite{DSS98,SRS08,Hastings09}.  As a result, very little 
has been known about the capacities of a quantum channel.

Lately there have been some surprising discoveries about the additivity properties of quantum capacities \cite{SST01,SST03,SY08,SS09,CH09,LWZG09}.  
A function on channels is called additive if its value on the tensor product of two channels is equal to the sum of the 
value on the individual channels: $f(\cN\ox\cM) = f(\cN) + f(\cM)$.  Additivity of a capacity means that the communication capabilities 
of channels don't interact when you use them together---a channel is good for the same amount of communication no matter what other channels are
available.  Conversely, when a capacity is nonadditive it means that the value of a channel for communication depends on what other channels 
it might be used with.  It was found in \cite{SY08} that the quantum capacity is strongly nonadditive.  In fact, there are pairs of channels
with $\cQ(\cN) = \cQ(\cM) = 0$ but $\cQ(\cN\ox \cM)>0$.  Something similar was found for the private capacity in \cite{LWZG09}, where channels
were presented with $\cP(\cN) = 0$ and $\cP(\cM) \approx 1$ but $\cP(\cN\ox \cM)\gg 1$.  So, it appears that the communication value of a 
quantum channel is not a simple function of the channel itself but also of the context in which it is used.

In this paper we present a family of channels displaying extensive
nonadditivity of private capacity, meaning additivity violations
proportional to the input size.  This involves two crucial innovations
over previous work.  First, our channels are much simpler than those
of \cite{SS09,BDSS06} and \cite{LWZG09} and do not rely on an
assumption of additivity of Holevo information as in
\cite{SS09,BDSS06}.  Second, while it was shown in \cite{LWZG09} that
$\cP$ is not additive, the violation is a vanishingly small fraction
of the channel's input size.  Our work shows additivity violation of
the strongest possible sort, with violations proportional to the log
of the input dimension.  Since $\log D$ is the largest possible capacity (classical or quantum)
for a channel with input dimension $D$, and thus the natural scale of the
capacity, violations of this sort
show that nonadditivity is an essential feature of the private
capacity.

\begin{figure}[htbp]
\includegraphics[width=3in]{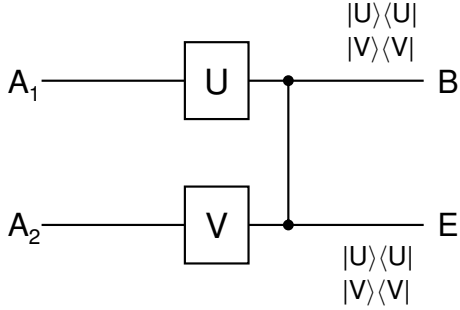}
\caption{Random Phase Coupling Channel. Unknown randomly chosen unitaries are applied independently to $A_1$ and $A_2$.  A controlled 
phase is then applied to between $A_1$ and $A_2$, which we now relabel $B$ and $E$, respectively.  The $E$ system is traced out, while the $B$
system is delivered to the receiver together with a classical description of the unitaries $U$ and $V$.  }
\label{Fig:Rocket}
\end{figure}

{\em Random Phase Coupling Channels---}
The channels we will focus on are pictured in FIG.~\ref{Fig:Rocket}.  $\cR_d$ has two $d$-dimensional 
inputs, $A_1$ and $A_2$.  After local random unitaries $U$ and $V$ are applied individually to these inputs, a controlled phase is applied 
and $A_2$ is discarded.  In addition to receiving the $A_1$ system (now relabeled $B$), the receiver is given a classical register describing
which $U$ and $V$ were chosen.

More formally, we let $W_{UV} = P U\ox V$,
where $P = \sum_{i,j}\omega^{ij}\proj{i}_{A_1}\ox \proj{j}_{A_2} $ is the controlled phase gate on $A_1A_2$ and $\omega$ is a primitive $d$th root 
of unity. Note that $W_{UV}$ maps $A_1A_2$ to $BE$.
We let $\cR_{UV}(\phi) = \Tr_E W_{UV}\phi W_{UV}^\dg$, and define our channel $\cR_d = \EE_{U,V} \cR_{UV}\ox\proj{U}\ox \proj{V}$, where $\EE_{U,V}$ is
the expectation with respect to random variables $U$ and $V$.  Throughout 
we will let $U^n = U_1\ox \dots \ox U_n$ and similarly for $V$, and define $\cR_{U^nV^n} = \cR_{U_1V_1}\ox \dots \ox \cR_{U_nV_n} $.
 
A key feature of our channel $\cR_d$ is that the unitaries $U$ and $V$ are unknown to the sender.  If she were told $V$ before using the channel,
she would be able to carefully choose the input to $A_2$ so as to avoid any dephasing of $A_1$.  However, since she does not, she is unable to avoid 
choosing an input that results in a significant amount of dephasing when averaged over the choice of $V$.  This rules out much quantum
capacity.  If at least she knew $U$ in 
advance, she'd still be able to send classical messages in the basis of dephasing, but since the dephasing is in a random basis, known only 
to the receiver even the classical capacity of the channel is low.

However, when entanglement between sender and receiver is available, things change dramatically.  We will see below that by having the sender
feed half of a maximally entangled pair into the $A_2$ system, the other half sitting with the receiver, the channel can transmit quantum
information at a rate of $\log d$.  Since quantum communication is necessarily private, private communication is also possible at this rate.  
We will be able to use this property, together with the probabilistic entanglement provided by a $50\%$ erasure channel, $\cA_d^e$, which itself
has zero private capacity, to show that $\cR_d\ox{\cA_d^e}$ has a quantum
capacity of at least $(1/2) \log d$, even though the individual private capacities are much smaller.

{\em Small Classical Capacity---}
Our goal now is to show that $\cR_d$ has a small classical capacity---$\cC(\cR_d)\leq 2$.  To do this, 
we first have to review some well known facts.

For an ensemble $\cE = \{p_i,\phi_i\}$, we define the Holevo information \cite{Holevo73}
\begin{equation}\nonumber
\chi(\cN,\cE) = S\left( \rho\right)-\sum_i p_i S\left(\rho_i\right)
\end{equation}
where $\rho_i = \cN(\phi_i)$ and $\rho = \sum_i p_i \rho_i$, $S(\rho) = -\Tr\rho \log \rho$, and throughout logarithms will
be taken base two.  The classical capacity of a quantum channel is given as follows \cite{Holevo98,SW97}:
\begin{equation}\label{Eq:ClassicalCapacity}
\cC(\cN) = \lim_{n\rightarrow \infty} \frac{1}{n}\max_\cE\chi\left(\cN^{\ox n},\cE\right).
\end{equation}

Our main technical result is the following lemma, whose proof may be skipped on a first reading.

\noindent{\bf Lemma 1} Let $\rho_{U^nV^n} = \cR_{U^nV^n}(\psi_{A_1^nA_2^n})$.  Then  for $d \geq 9$
\begin{equation}\nonumber
\EE_{U^nV^n} S(\rho_{U^nV^n}) \geq n(\log d - 2).
\end{equation}
 {\bf Proof} First note that $\EE S(\rho_{U^nV^n}) \geq -\log \EE \Tr\rho^2_{U^nV^n}$,
so it will suffice to give an upper bound on $\EE \Tr\rho^2_{U^nV^n}$.  To do this, we'll use the fact
that 
\begin{equation}\nonumber
\Tr\left(\rho^2_{U^nV^n}\right) = \Tr(\rho^{B^n}_{U^nV^n} \ox \rho^{{B^\prime}^n}_{U^nV^n} F_{B^n{B^\prime}^n})
\end{equation}
where $F_{B^n{B^\prime}^n}$ is the unitary that
swaps $B^n$ and ${B^\prime}^n$.  In particular, letting $X = \left(P_n^\dg \ox P_n^\dg\right) 
F_{B^n{B^\prime}^n}\ox I_{E^n{E^\prime}^n} \left(P_n \ox P_n\right)$ we find that 
\begin{equation}\nonumber
\EE\Tr(\rho_{U^nV^n}^2) = \Tr\left( \Psi X \right)
\end{equation}
where $\Psi = \EE U^n_{BB'}\ox V^n_{EE'} \psi_{A^n_1A^n_2}\ox \psi_{{A'}^n_1{A'}^n_2}{U^\dg}^n_{BB'}\ox {V^\dg}^n_{EE'}$ with 
$U^n_{BB'}= \ox_{l=1}^{n}(U_l \ox U_l)$, similarly for $V^n_{EE'}$
and $P_n = P^{\ox n}$.  In fact, we won't even need to calculate $\Psi$ exactly, since by Schur's Lemma \cite{Simon96} it takes the form
\begin{equation}\label{Eq:IrrepM}
\Psi = \sum_{{\mathbf s}_b{\mathbf s}_e} \alpha_{{\mathbf s}_b{\mathbf s}_e}\frac{\Pi^{BB'}_{{\mathbf s}_b}}{d^B_{{\mathbf s}_b}} \ox \frac{\Pi^{EE'}_{{\mathbf s}_e}}{d^E_{{\mathbf s}_e}},
\end{equation}
where ${\mathbf s}_b$ and ${\mathbf s}_e$ are $n$-bit strings,  and $\alpha_{{\mathbf s}_b{\mathbf s}_e}$ are probabilities.
Here we have used the notation
\begin{equation}\nonumber
\Pi^{BB'}_{{\mathbf s}_b} = \Pi^{B_1B'_1}_{({\mathbf s}_b)_1}\otimes \dots \otimes \Pi^{B_nB'_n}_{({\mathbf s}_b)_n},
\end{equation}
where $\Pi^{BB'}_0$ is the projector onto the symmetric space of $BB'$, $\Pi^{BB'}_1$ projects onto the antisymmetric space, similarly 
for $\Pi^{EE'}_{{\mathbf s}_e}$, and $d^B_{{\mathbf s}_b}$ and $d^E_{{\mathbf s}_e}$ are the ranks of 
$\Pi^{BB'}_{{\mathbf s}_b}$ and $\Pi^{EE'}_{{\mathbf s}_e}$, respectively.

Because of Eq.~(\ref{Eq:IrrepM}), we can understand $\Tr(\Psi X)$ by focusing on a term of the form
\begin{equation}\nonumber
\Tr\left(\Pi^{BB'}_{{\mathbf s}_b}\ox\Pi^{EE'}_{{\mathbf s}_e}X\right).
\end{equation}
This, in turn, is the product of $n$ terms of the form
\begin{equation}\nonumber
\Tr \left( \Pi^{BB'}_{s_b}\ox\Pi^{EE'}_{s_e} (P_{BE}\ox P_{B'E'})I\ox F_{BB'}(P_{BE}\ox P_{B'E'})^\dg\right),
\end{equation}
which it is easy to verify equals
\begin{equation}\nonumber
\frac{(-1)^{s_e}}{4}\sum_{i,j,i'j'}\omega^{(i-i')(j-j')}\left( \delta_{ii'}+ (-1)^{s_b}\right)\left(\delta_{jj'}+(-1)^{s_e} \right).
\end{equation}
  Evaluating this sum explicitly gives us
\begin{equation}\nonumber
\frac{(-1)^{s_e}}{4}\left( d^2 + (-1)^{s_b}d^3 + (-1)^{s_e}d^3 + (-1)^{s_b+s_e}d^3\right)
\end{equation}
which, in turn, is no larger than $d^2(3d+1)/4$.  As a result, we have
\begin{equation}\nonumber
\Tr(\Pi^{BB'}_{{\mathbf s}_b}\ox\Pi^{EE'}_{{\mathbf s}_e}X) \leq (d^2(3d+1)/4)^n.
\end{equation}
Using this bound in combination with the fact that $d^B_{{\mathbf s}_b} \geq (d(d-1)/2)^n$ and similarly for $d^E_{{\mathbf s}_e}$, 
we find
\begin{equation}\nonumber
\Tr( \Psi X) \leq \frac{(d^2(3d+1)/4)^n}{(d(d-1)/2)^{2n}} = \left( \frac{3d+1}{(d-1)^2}\right)^{n}.
\end{equation}

Finally, we translate this back to a lower bound on the average entropy of $B^n$:
\begin{equation}\nonumber
\EE S(\rho_{U^nV^n}) \geq n \log\left( (d-1)^2/(3d+1)\right),
\end{equation}
which, noting that for $d \geq 9$ we have $(d-1)^2/(3d+1) \geq d/4$ proves the result. \hfill $\Box$

We now turn to the classical capacity of our channel.  
Because our channels have infinite dimensional classical registers, to avoid technical complications we
write the Holevo quantity for $\cR^{\ox n}$ together with ensemble $\cE$ as

\begin{equation}\nonumber
\chi(\cR^{\ox n},\cE) = \EE_{U^nV^n}  \chi\left(\cR_{U^nV^n},\cE\right). 
\end{equation}

Now, for any input ensemble $\{p_i,\phi_i\}$ to $n$ copies of our channel $\cR_d^{\ox n}$, we have 
\begin{equation}\nonumber
S\left(\cR_{U^nV^n}\left(\sum_i p_i \phi_i\right)\right)\leq n \log d.
\end{equation}
Furthermore, by the Lemma, for each $\phi_i$, the entropy of $\cR_{U^nV^n}(\phi_i)$ averaged over $U^nV^n$ is at least $n(\log d-2)$.  As a result, 
for any ensemble $\cE$, we have $\EE\chi(\cR_{U^nV^n},\cE) \leq 2n$.  In light of Eq.~(\ref{Eq:ClassicalCapacity}) this gives $\cC(\cR_d) \leq 2$.

{\em Large Joint Quantum Capacity---}

We now show that the joint quantum capacity of a random phase coupling channel, $\cR_d$, and a $50\%$ erasure channel
is at least $(1/2)\log d$.  To do this, we will need the following lower bound for the quantum capacity \cite{Lloyd97,Shor02,D03}, called the coherent information:
\begin{equation}\nonumber
\cQ(\cN) \geq \max_{\phi_{AA'}} \left( S(B) -S(AB)\right),
\end{equation}
where the entropies are evaluated on the state $(I\ox \cN)(\phi)$.  In our case, since $\cR_d$ has infinite dimensional classical outputs,
the correct lower bound to consider is the coherent information of the channel given  $U$ and $V$, averaged over $U,V$.

\begin{figure}[htbp]
\includegraphics[width=3in]{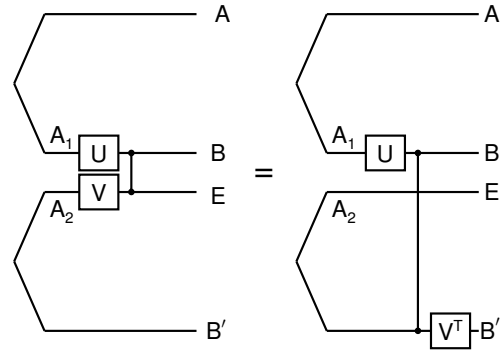}
\caption{Reversing random phase coupling with entanglement.  Using a maximally entangled state, $\ket{\phi_d}_{A_2B'}$, the action of $\cR_d$
on $A_1$ can be reversed.  This depends on the fact that for any $M$, $M\ox I\ket{\phi_d} = I\ox M^{T}\ket{\phi_d}$, so that by inserting half of
$\ket{\phi_d}_{A_2B'}$ into $A_2$ the receiver holding $B$ and $B'$ can invert $U$,$V^T$ and $P$, the controlled phase operation.  
By feeding $B'$ into a $50\%$ erasure channel, 
half the time this gives a coherent information of $\log d$ between sender and receiver.  The other half of the time the coherent information is
exactly zero so that the overall coherent information is $(1/2)\log d$.}
\label{Fig:UR}
\end{figure}

The way to use the two channels together is as follows.  We prepare two maximally entangled states $\ket{\varphi}= \ket{\phi_d}_{AA_1}\ket{\phi_d}_{B'A_2}$ and
feed $A_1A_2$ into $\cR_d$ and $B'$ into ${\cA_d^e}$.  The coherent information then breaks up into a sum of two terms.  The first, which occurs
when the input to ${\cA_d^e}$ is not erased (which has probability $1/2$) is equal to $\log d$, as explained in FIG.~\ref{Fig:UR}.  The second, which occurs when ${\cA_d^e}$
emits an erasure flag (and also has probability $1/2$), is the coherent information of a completely dephasing channel in a basis known only to the receiver.  The resulting coherent 
information in this second case is exactly zero.  The coherent information of $\cR_d\ox {\cA_d^e}$ evaluated on $\ket{\varphi}$ is just the average of 
these two, $(1/2)\log d$.  Recalling that $\cP(\cR_d) \leq \cC(\cR_d) \leq 2$ and $\cP(\cA_d^e)=0$ gives the nonadditivity we sought.

{\em Discussion---}We have shown that the quantum and private
capacities of a quantum channel are extremely nonadditive.  This
nonadditivity illustrates, in contrast to the classical theory, that
the communication capabilities of a quantum channel depend inherently
on the setting in which it is used.  Our construction is essentially a
simplification and strengthening of the retrocorrectible channels
studied in \cite{BDSS06,SS09}.  As a result, in addition to
nonadditivity, our channels also provide unconditional separations of
capacities which were only conjectured in \cite{BDSS06}.  

In particular we can show, contrary to the classical case, that the
classical capacity of a quantum channel, assisted by backwards
classical communication, may substantially exceed the unassisted
capacity.  To see this, note that if, upon putting halves of
maximally entangled states into $\cR_d$, the receiver tells the sender 
$U$ and $V$, she can easily invert $U$, $V$, and $P$ to establish a $d$-dimensional
maximally entangled state.  In fact, we will use two copies of $\cR_d$ to establish two $d$-dimensional 
maximally entangled states.  We will use one maximally entangled state, together with a third copy of 
the channel to simulate a perfect d-dimensional quantum channel as explained in FIG.\ref{Fig:UR}.  The second 
maximally entangled state can be used, together with this perfect channel to send $2\log d$ classical bits
by using superdense coding \cite{BW92}.  This results in a back-assisted classical capacity of at least $2/3 \log d$
with an unassisted classical capacity of no more than 2.

In terms of magnitude, superadditivity of private and quantum
capacities cannot exceed $\log D$ for channels with input dimension
$D$.  Our channels achieve $(1/4)\log D$ in the limit of large input
dimension.  We suspect this is optimal both because of the simplicity
of these channels and the structure of all known examples of
superadditivity.  However, we have not yet found a proof.

While the channels above have finite input dimension, as described
they have infinite dimensional (indeed, continuous) outputs.  This is
not a serious drawback, because our main technical argument (Lemma 1)
depends only on the fact that a random unitary ensemble is a so-called
two-design.  Luckily, the Clifford group is finite and has this
property \cite{DLT01}, so that we can replace the infinite output
above with an output of size $O((\log d)^2)$.

The superactivation effect of \cite{SY08} is not yet completely
understood.  From that work it appeared that the fundamental effect
was one of transforming noncoherent privacy\cite{HHHO03,HPHH05} to
coherent communication with the assistance of an erasure channel.
However, our results here show conclusively that strong superadditivity
of this sort is possible using channels with almost no private capacity, and
in fact the private classical capacity is just as superadditive.  What
exactly is the source of superadditivity and which channels can be
activated remain elusive open questions.

{\em Acknowledgements---} We are grateful to Ke Li and Andreas Winter for providing us an early draft of \cite{LWZG09}.   We both 
received support from the DARPA QUEST program 
under contract no. HR0011-09-C-0047.


\begin{thebibliography}{24}
\expandafter\ifx\csname natexlab\endcsname\relax\def\natexlab#1{#1}\fi
\expandafter\ifx\csname bibnamefont\endcsname\relax
  \def\bibnamefont#1{#1}\fi
\expandafter\ifx\csname bibfnamefont\endcsname\relax
  \def\bibfnamefont#1{#1}\fi
\expandafter\ifx\csname citenamefont\endcsname\relax
  \def\citenamefont#1{#1}\fi
\expandafter\ifx\csname url\endcsname\relax
  \def\url#1{\texttt{#1}}\fi
\expandafter\ifx\csname urlprefix\endcsname\relax\def\urlprefix{URL }\fi
\providecommand{\bibinfo}[2]{#2}
\providecommand{\eprint}[2][]{\url{#2}}

\bibitem[{\citenamefont{Shannon}(1948)}]{Shannon48}
\bibinfo{author}{\bibfnamefont{C.~E.} \bibnamefont{Shannon}},
  \bibinfo{journal}{Bell Syst. Tech. J.} \textbf{\bibinfo{volume}{27}},
  \bibinfo{pages}{379} (\bibinfo{year}{1948}).

\bibitem[{\citenamefont{Richardson and Urbanke}(2003)}]{RU03}
\bibinfo{author}{\bibfnamefont{T.}~\bibnamefont{Richardson}} \bibnamefont{and}
  \bibinfo{author}{\bibfnamefont{R.}~\bibnamefont{Urbanke}},
  \bibinfo{journal}{IEEE Communications Magazine}
  \textbf{\bibinfo{volume}{41}}, \bibinfo{pages}{126} (\bibinfo{year}{2003}).

\bibitem[{\citenamefont{Devetak}(2005)}]{D03}
\bibinfo{author}{\bibfnamefont{I.}~\bibnamefont{Devetak}},
  \bibinfo{journal}{IEEE Trans. Inf. Theory} \textbf{\bibinfo{volume}{51}},
  \bibinfo{pages}{44} (\bibinfo{year}{2005}),
  \bibinfo{note}{ar{X}iv:quant-ph/0304127}.

\bibitem[{\citenamefont{Bennett and Brassard}(1984)}]{BB84}
\bibinfo{author}{\bibfnamefont{C.~H.} \bibnamefont{Bennett}} \bibnamefont{and}
  \bibinfo{author}{\bibfnamefont{G.}~\bibnamefont{Brassard}},
  \bibinfo{journal}{Proceedings of the IEEE International Conference on
  Computers, Systems and Signal Processing} p. \bibinfo{pages}{175}
  (\bibinfo{year}{1984}).

\bibitem[{\citenamefont{Di{V}incenzo et~al.}(1998)\citenamefont{Di{V}incenzo,
  Shor, and Smolin}}]{DSS98}
\bibinfo{author}{\bibfnamefont{D.}~\bibnamefont{Di{V}incenzo}},
  \bibinfo{author}{\bibfnamefont{P.~W.} \bibnamefont{Shor}}, \bibnamefont{and}
  \bibinfo{author}{\bibfnamefont{J.~A.} \bibnamefont{Smolin}},
  \bibinfo{journal}{Phys. Rev. A} \textbf{\bibinfo{volume}{57}},
  \bibinfo{pages}{830} (\bibinfo{year}{1998}),
  \bibinfo{note}{ar{X}iv:quant-ph/9706061}.

\bibitem[{\citenamefont{Smith et~al.}(2008)\citenamefont{Smith, Renes, and
  Smolin}}]{SRS08}
\bibinfo{author}{\bibfnamefont{G.}~\bibnamefont{Smith}},
  \bibinfo{author}{\bibfnamefont{J.}~\bibnamefont{Renes}}, \bibnamefont{and}
  \bibinfo{author}{\bibfnamefont{J.~A.} \bibnamefont{Smolin}},
  \bibinfo{journal}{Phys. Rev. Lett.} \textbf{\bibinfo{volume}{100}},
  \bibinfo{pages}{170502} (\bibinfo{year}{2008}),
  \bibinfo{note}{ar{X}iv:quant-ph/0607018}.

\bibitem[{\citenamefont{Hastings}(2009)}]{Hastings09}
\bibinfo{author}{\bibfnamefont{M.}~\bibnamefont{Hastings}},
  \bibinfo{journal}{Nat. Phys.} \textbf{\bibinfo{volume}{5}},
  \bibinfo{pages}{255} (\bibinfo{year}{2009}),
  \bibinfo{note}{ar{X}iv:0809.3972}.

\bibitem[{\citenamefont{Shor et~al.}(2001)\citenamefont{Shor, Smolin, and
  Terhal}}]{SST01}
\bibinfo{author}{\bibfnamefont{P.}~\bibnamefont{Shor}},
  \bibinfo{author}{\bibfnamefont{J.}~\bibnamefont{Smolin}}, \bibnamefont{and}
  \bibinfo{author}{\bibfnamefont{B.}~\bibnamefont{Terhal}},
  \bibinfo{journal}{Phys. Rev. Lett.} \textbf{\bibinfo{volume}{86}},
  \bibinfo{pages}{2681} (\bibinfo{year}{2001}).

\bibitem[{\citenamefont{Shor et~al.}(2003)\citenamefont{Shor, Smolin, and
  Thapliyal}}]{SST03}
\bibinfo{author}{\bibfnamefont{P.}~\bibnamefont{Shor}},
  \bibinfo{author}{\bibfnamefont{J.}~\bibnamefont{Smolin}}, \bibnamefont{and}
  \bibinfo{author}{\bibfnamefont{A.}~\bibnamefont{Thapliyal}},
  \bibinfo{journal}{Phys. Rev. Lett.} \textbf{\bibinfo{volume}{90}},
  \bibinfo{pages}{107901} (\bibinfo{year}{2003}).

\bibitem[{\citenamefont{Smith and Yard}(2008)}]{SY08}
\bibinfo{author}{\bibfnamefont{G.}~\bibnamefont{Smith}} \bibnamefont{and}
  \bibinfo{author}{\bibfnamefont{J.}~\bibnamefont{Yard}},
  \bibinfo{journal}{Science} \textbf{\bibinfo{volume}{321}},
  \bibinfo{pages}{1812} (\bibinfo{year}{2008}),
  \bibinfo{note}{arXiv:0807.4935}.

\bibitem[{\citenamefont{Smith and Smolin}(2009)}]{SS09}
\bibinfo{author}{\bibfnamefont{G.}~\bibnamefont{Smith}} \bibnamefont{and}
  \bibinfo{author}{\bibfnamefont{J.~A.} \bibnamefont{Smolin}},
  \bibinfo{journal}{Phys. Rev. Lett.} \textbf{\bibinfo{volume}{102}},
  \bibinfo{pages}{010501} (\bibinfo{year}{2009}),
  \bibinfo{note}{ar{X}iv:0810.0276}.

\bibitem[{\citenamefont{Czekaj and Horodecki}(2009)}]{CH09}
\bibinfo{author}{\bibfnamefont{L.}~\bibnamefont{Czekaj}} \bibnamefont{and}
  \bibinfo{author}{\bibfnamefont{P.}~\bibnamefont{Horodecki}},
  \bibinfo{journal}{Phys. Rev. Lett.} \textbf{\bibinfo{volume}{102}},
  \bibinfo{pages}{110505} (\bibinfo{year}{2009}),
  \bibinfo{note}{ar{X}iv:0807.3977}.

\bibitem[{\citenamefont{Li et~al.}()\citenamefont{Li, Winter, Zou, and
  Guo}}]{LWZG09}
\bibinfo{author}{\bibfnamefont{K.}~\bibnamefont{Li}},
  \bibinfo{author}{\bibfnamefont{A.}~\bibnamefont{Winter}},
  \bibinfo{author}{\bibfnamefont{X.}~\bibnamefont{Zou}}, \bibnamefont{and}
  \bibinfo{author}{\bibfnamefont{G.}~\bibnamefont{Guo}},
  \bibinfo{note}{ar{X}iv:0903.4308}.

\bibitem[{\citenamefont{Bennett et~al.}(2006)\citenamefont{Bennett, Devetak,
  Shor, and Smolin}}]{BDSS06}
\bibinfo{author}{\bibfnamefont{C.~H.} \bibnamefont{Bennett}},
  \bibinfo{author}{\bibfnamefont{I.}~\bibnamefont{Devetak}},
  \bibinfo{author}{\bibfnamefont{P.~W.} \bibnamefont{Shor}}, \bibnamefont{and}
  \bibinfo{author}{\bibfnamefont{J.~A.} \bibnamefont{Smolin}},
  \bibinfo{journal}{Phys. Rev. Lett.} \textbf{\bibinfo{volume}{96}},
  \bibinfo{pages}{150502} (\bibinfo{year}{2006}),
  \bibinfo{note}{ar{X}iv:quant-ph/0406086}.

\bibitem[{\citenamefont{Holevo}(1973)}]{Holevo73}
\bibinfo{author}{\bibfnamefont{A.~S.} \bibnamefont{Holevo}}, in
  \emph{\bibinfo{booktitle}{Proceedings of the second {Japan-USSR} Symposium on
  Probability Theory}}, edited by
  \bibinfo{editor}{\bibfnamefont{G.}~\bibnamefont{Maruyama}} \bibnamefont{and}
  \bibinfo{editor}{\bibfnamefont{J.~V.} \bibnamefont{Prokhorov}}
  (\bibinfo{publisher}{Springer-Verlag}, \bibinfo{address}{Berlin},
  \bibinfo{year}{1973}), vol. \bibinfo{volume}{330} of
  \emph{\bibinfo{series}{Lecture Notes in Mathematics}}, pp.
  \bibinfo{pages}{104--119}.

\bibitem[{\citenamefont{Holevo}(1998)}]{Holevo98}
\bibinfo{author}{\bibfnamefont{A.}~\bibnamefont{Holevo}},
  \bibinfo{journal}{IEEE Trans. Inform. Theory} \textbf{\bibinfo{volume}{44}},
  \bibinfo{pages}{269} (\bibinfo{year}{1998}).

\bibitem[{\citenamefont{Schumacher and Westmoreland}(1997)}]{SW97}
\bibinfo{author}{\bibfnamefont{B.}~\bibnamefont{Schumacher}} \bibnamefont{and}
  \bibinfo{author}{\bibfnamefont{M.}~\bibnamefont{Westmoreland}},
  \bibinfo{journal}{Phys. Rev. A} \textbf{\bibinfo{volume}{56}},
  \bibinfo{pages}{131} (\bibinfo{year}{1997}).

\bibitem[{\citenamefont{Simon}(1996)}]{Simon96}
\bibinfo{author}{\bibfnamefont{B.}~\bibnamefont{Simon}},
  \emph{\bibinfo{title}{Representations of Finite and Compact Groups}}
  (\bibinfo{publisher}{Amer. Math. Soc.}, \bibinfo{year}{1996}).

\bibitem[{\citenamefont{Shor}()}]{Shor02}
\bibinfo{author}{\bibfnamefont{P.~W.} \bibnamefont{Shor}},
  \bibinfo{note}{lecture notes, MSRI Workshop on Quantum Computation, 2002.
  Available online at http://www.msri.org/publications/ln/msri/2002/\\
  quantumcrypto/shor/1/}.

\bibitem[{\citenamefont{Lloyd}(1997)}]{Lloyd97}
\bibinfo{author}{\bibfnamefont{S.}~\bibnamefont{Lloyd}},
  \bibinfo{journal}{Phys. Rev. A} \textbf{\bibinfo{volume}{55}},
  \bibinfo{pages}{1613} (\bibinfo{year}{1997}).

\bibitem[{\citenamefont{Charles H.~Bennett}(1992)}]{BW92}
\bibinfo{author}{\bibfnamefont{S.~J.~W.} \bibnamefont{Charles H.~Bennett}},
  \bibinfo{journal}{\prl} \textbf{\bibinfo{volume}{69}}, \bibinfo{pages}{2881 }
  (\bibinfo{year}{1992}).

\bibitem[{\citenamefont{Divincenzo et~al.}(2002)\citenamefont{Divincenzo,
  Leung, and Terhal}}]{DLT01}
\bibinfo{author}{\bibfnamefont{D.~P.} \bibnamefont{Divincenzo}},
  \bibinfo{author}{\bibfnamefont{D.~W.} \bibnamefont{Leung}}, \bibnamefont{and}
  \bibinfo{author}{\bibfnamefont{B.~M.} \bibnamefont{Terhal}},
  \bibinfo{journal}{IEEE Trans. Info. Theory} \textbf{\bibinfo{volume}{48}},
  \bibinfo{pages}{580} (\bibinfo{year}{2002}),
  \bibinfo{note}{ar{X}iv:quant-ph/0103098}.

\bibitem[{\citenamefont{Horodecki et~al.}(2005)\citenamefont{Horodecki,
  Horodecki, Horodecki, and Oppenheim}}]{HHHO03}
\bibinfo{author}{\bibfnamefont{K.}~\bibnamefont{Horodecki}},
  \bibinfo{author}{\bibfnamefont{M.}~\bibnamefont{Horodecki}},
  \bibinfo{author}{\bibfnamefont{P.}~\bibnamefont{Horodecki}},
  \bibnamefont{and}
  \bibinfo{author}{\bibfnamefont{J.}~\bibnamefont{Oppenheim}},
  \bibinfo{journal}{Phys. Rev. Lett.} \textbf{\bibinfo{volume}{94}},
  \bibinfo{pages}{160502} (\bibinfo{year}{2005}).

\bibitem[{\citenamefont{Horodecki et~al.}(2008)\citenamefont{Horodecki,
  Pankowski, Horodecki, and Horodecki}}]{HPHH05}
\bibinfo{author}{\bibfnamefont{K.}~\bibnamefont{Horodecki}},
  \bibinfo{author}{\bibfnamefont{L.}~\bibnamefont{Pankowski}},
  \bibinfo{author}{\bibfnamefont{M.}~\bibnamefont{Horodecki}},
  \bibnamefont{and}
  \bibinfo{author}{\bibfnamefont{P.}~\bibnamefont{Horodecki}},
  \bibinfo{journal}{IEEE Trans. Info. Theory} \textbf{\bibinfo{volume}{54}},
  \bibinfo{pages}{2621} (\bibinfo{year}{2008}).

\end{thebibliography}
\end{document}